\documentclass{iopconfser}
\usepackage{amsmath}
\usepackage{orcidlink}
\usepackage{xcolor}
\usepackage{mathrsfs}
\usepackage{subcaption}

\begin{document}

\title{PINNGraPE: Physics Informed Neural Network for Gravitational wave Parameter Estimation}

\author{Leigh Smith$^{1,2}$\orcidlink{0000-0002-3035-0947}, Matteo Scialpi$^{3,4}$\orcidlink{0009-0007-6434-1460}, Francesco Di Clemente$^{5}$\orcidlink{0000-0002-8257-3819} and Micha\l~Bejger$^{4,6}\orcidlink{0000-0002-4991-8213}$}

\affil{$^1$Dipartimento di Fisica, Università di Trieste, Via Valerio 2, 34127 Trieste, Italy}
\affil{$^2$INFN, Sezione di Trieste, I-34127 Trieste, Italy}
\affil{$^3$Dipartimento di Fisica e Scienze della Terra, Università di Ferrara, Via Saragat 1, 44122 Ferrara, Italy}
\affil{$^4$INFN Sezione di Ferrara, Via Saragat 1, 44122 Ferrara, Italy}
\affil{$^5$Department of Physics, University of Houston, Houston, TX 77204, USA}
\affil{$^6$Nicolaus Copernicus Astronomical Center, Polish Academy of Sciences, Bartycka 18, 00-716, Warsaw, Poland}

\email{leighann.smith@units.it}

\begin{abstract}

Weakly-modelled searches for gravitational waves are essential for ensuring that all potential sources are accounted for in detection efforts, as they make minimal assumptions regarding source morphology. While these searches primarily target generic transient sources, they are also highly effective at identifying a broad range of compact binary coalescences, demonstrated by the weakly-modelled search algorithm coherent WaveBurst being the first to detect GW150914. Despite their ability to detect compact binaries with diverse properties, the accurate estimation of source parameters from their output remains to be a challenging task. To overcome this, we leverage physics-informed neural networks, which serve as a powerful tool for parameter estimation by applying physical constraints through the universal differential equation governing a compact binary system. With this approach, we rapidly infer the mass parameters of binary black hole merger systems to within 7\% from only the time-frequency representation of the gravitational wave signal. 
\end{abstract}

\section{Introduction}

coherent WaveBurst (cWB) \cite{Klimenko_2008cwb,Klimenko:2015cwb,Drago:2020cwb} is a weakly-modelled gravitational wave (GW) detection algorithm which relies on excess coherent energy across a network of detectors. While it is used extensively in searches for generic GW transients which cannot rely on well-modelled waveforms, it is also highly efficient in detecting binary black hole (BBH) events without assuming a specific source waveform \cite{cwb-150914,cwb-190521,cwb-bbh-o3}. 
Despite this, it lacks the ability to accurately infer source parameters from the reconstructed waveform, aside from sky location and an estimate of chirp mass from the likelihood. 
To overcome this problem we utilise physics-informed neural networks (PINNs), which aim to learn the behaviour of physical systems by embedding governing differential equations and physical relations in the loss function of a neural network \cite{Raissi:2017pinn}. 
In this work, we train a PINN to learn the governing equations of a coalescing binary, with the goal of inferring mass properties of a BBH from only the GW signal. We consider the 1.5 post-Newtonian (PN) frequency evolution of the merging binary \cite{Cutler:1994}: 

\begin{equation}
    \frac{df}{dt}=\frac{96}{5}\pi^{8/3}\left(\frac{GM_\odot}{c^3}\frac{\mathcal{M}}{M_\odot}\right)^{5/3}f^{11/3}\left[1-\left(\frac{743}{336}+\frac{11}{4}\eta\right)\varepsilon^{2/3}+4\pi\varepsilon\right], \qquad \varepsilon=\frac{GM_\odot}{c^3}\left(\frac{M_\mathrm{tot}}{M_\odot}\right)\pi f
    \label{eq:df_dt_1hPN}
\end{equation}
for chirp mass $\mathcal{M}$, total mass $M_\mathrm{tot}$ and symmetric mass ratio $\eta$. Expanding to further PN orders increases the parameters in this relation, however for this study we consider only mass properties. These mass parameters are related to the component masses of the system through the following 3 equations:

\begin{equation}
    \mathcal{M}=\frac{\left(m_\textrm{1}m_\textrm{2}\right)^{3/5}}{\left(m_\textrm{1}+m_\textrm{2}\right)^{1/5}}\,,\qquad M_\mathrm{tot}=m_\textrm{1}+m_\textrm{2}\,,\qquad\eta=\frac{m_\textrm{1}m_\textrm{2}}{(m_\textrm{1}+m_\textrm{2})^2}\,,
    \label{eq:masses}
\end{equation}
which can be re-arranged to form a direct relation between $\mathcal{M}$, $M_\mathrm{tot}$ and $\eta$:
\begin{equation}\label{eq:mass_relation}
    \mathcal{M} = M_\mathrm{tot}\eta^{3/5} 
\end{equation}
From the above equations we are able to construct a neural network that will learn the frequency evolution of the GW signal through physics-informed losses, thus outputting the related mass properties.

\section{Data and model architecture}
For a detected a signal, cWB constructs likelihood maps in the time-frequency (TF) domain, over 7 different TF resolutions. The aim is to eventually use the TF outputs directly from cWB, however for this preliminary study we construct TF data to mimic the structure of cWB outputs using the PyCBC package \cite{pycbc}. This data consists of BBH simulations injected into Gaussian noise coloured by the design sensitivity power spectral density of the Advanced LIGO (aLIGO) detector \cite{aLIGO}. Each injection is transformed into the TF domain over 7 resolutions. 
In total we create 10,000 signals, varied over distance [500,1000] Mpc and sampled uniformly from total mass [30,100] $M_{\odot}$ and symmetric mass ratio [0.1,0.25]. The remaining mass properties are calculated via Eqs. \ref{eq:masses} and \ref{eq:mass_relation}.
The data is divided into 70\% training, 20\% validation and 10\% testing data. 

\begin{figure}
    \centering
    \includegraphics[width=\linewidth]{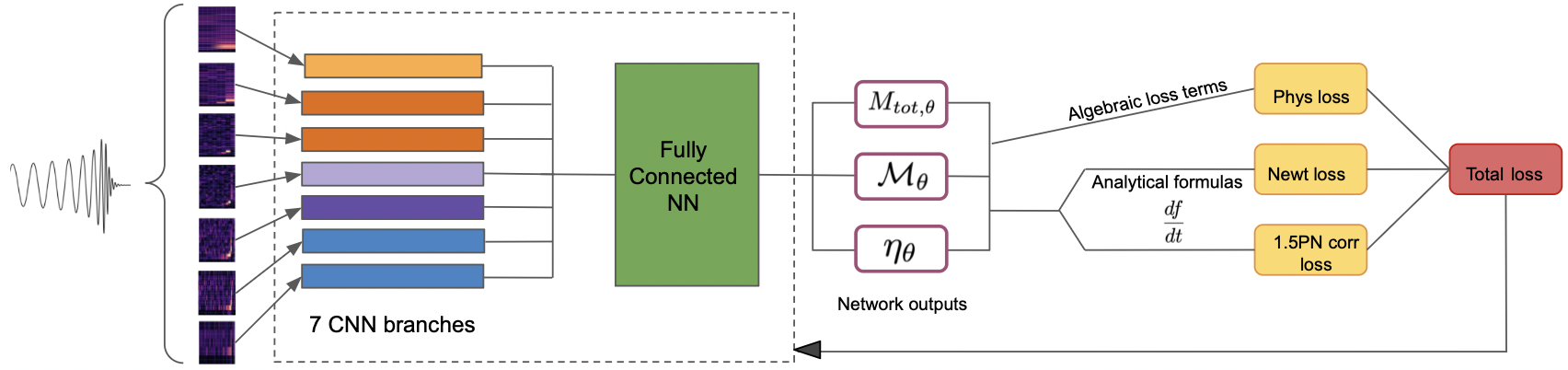}
    \caption{Diagram of the multi-branch CNN architecture. 7 TF representations are input into the network, each processed by a different CNN branch. CNN outputs are flattened and inserted into a fully-connected neural network, which directly outputs 3 mass parameters. Physical constraints are considered in the loss terms, used to update the hyper-parameters of the network. 
    }
    \label{fig:NN}
\end{figure}

We employ a multi-branch convolutional neural network (CNN) as the base of our model, an overview of which is seen in Fig. \ref{fig:NN}. Each of the 7 TF resolutions is input into a separate CNN branch, designed to handle the varied dimensionality of the inputs with different convolutional and maxpooling layer definitions. 
The outputs of the CNN branches are flattened into a fully-connected neural network, which has 3 output neurons corresponding to $\{M_\mathrm{tot},\mathcal{M}, \eta\}_{\theta}$.
While only 2 output mass parameters are needed to infer all mass properties theoretically, the inclusion of all 3 parameters in Eq. \ref{eq:mass_relation} allows the network to learn the physical redundancies within this relation. This is enforced through the loss equation and ultimately improves the accuracy of the network.
The physics-informed loss is constructed from 2 main components. The first component is a mean squared loss between output and target terms of the analytical $df/dt$ formula, where output terms are calculated via $\{M_\mathrm{tot},\mathcal{M}, \eta\}_{\theta}$ and Eq. \ref{eq:df_dt_1hPN}. The $df/dt$ loss component is split by Newtonian and 1.5 PN correction terms due to the extreme magnitude difference in their values. The second component consists of auxiliary algebraic loss terms between the 3 outputs of the network. These are the terms that exploit the physical redundancy within Eq. \ref{eq:mass_relation}, which helps with the convergence and accuracy of the network. With the above considerations, the total loss is given by: 
\begin{multline}\label{eq:loss}
    \mathscr{L} = \mathscr{L}_\mathrm{Newt} + \mathscr{L}_\mathrm{1.5PN} + \mathscr{L}_\mathrm{Phys} = \frac{\beta_1}{B}\sum^{B}_{i=1}\Bigg(\frac{df_{\theta}}{dt} - \frac{df_T}{dt} \Bigg)^{2}_\mathrm{Newt} + \frac{\beta_2}{B}\sum^{B}_{i=1}\Bigg(\frac{df_{\theta}}{dt} - \frac{df_T}{dt} \Bigg)^{2}_\mathrm{1.5PN} \\
    + \frac{\beta_3}{B}\sum^{B}_{i=1}\Bigg(M_\mathrm{tot,\theta} \eta^{3/5}_{\theta} - \mathcal{M}_{\theta} \Bigg)^{2} + \frac{\beta_4}{B}\sum^{B}_{i=1}\Bigg(\mathcal{M}_{\theta} \eta^{-3/5}_{\theta} - M_\mathrm{tot, \theta} \Bigg)^{2} + \frac{\beta_5}{B}\sum^{B}_{i=1}\Bigg(\Bigg(\frac{\mathcal{M}_{\theta}}{M_\mathrm{tot, \theta}}\Bigg)^{3/5} - \eta_{\theta} \Bigg)^{2}
\end{multline}
where $B$ is the batch size and $\beta_1=10^{-6}, \beta_2=10^4, \beta_3=\beta_4=\beta_5=1 $ are weighting factors. Values related to the network outputs are notated by $\theta$, while target values are notated by $T$. The subscript $i$ has been dropped from all values for simplicity. 
In total the network takes $\sim 80$ minutes to train on a GPU, with a \textsc{ReduceLROnPlateau} scheduler decreasing the learning rate to $< 10^{-6}$.

\section{Discussion and future outlook}

\begin{figure}
    \centering
     \begin{subfigure}{0.49\textwidth}
         \centering
         \includegraphics[width=\textwidth]{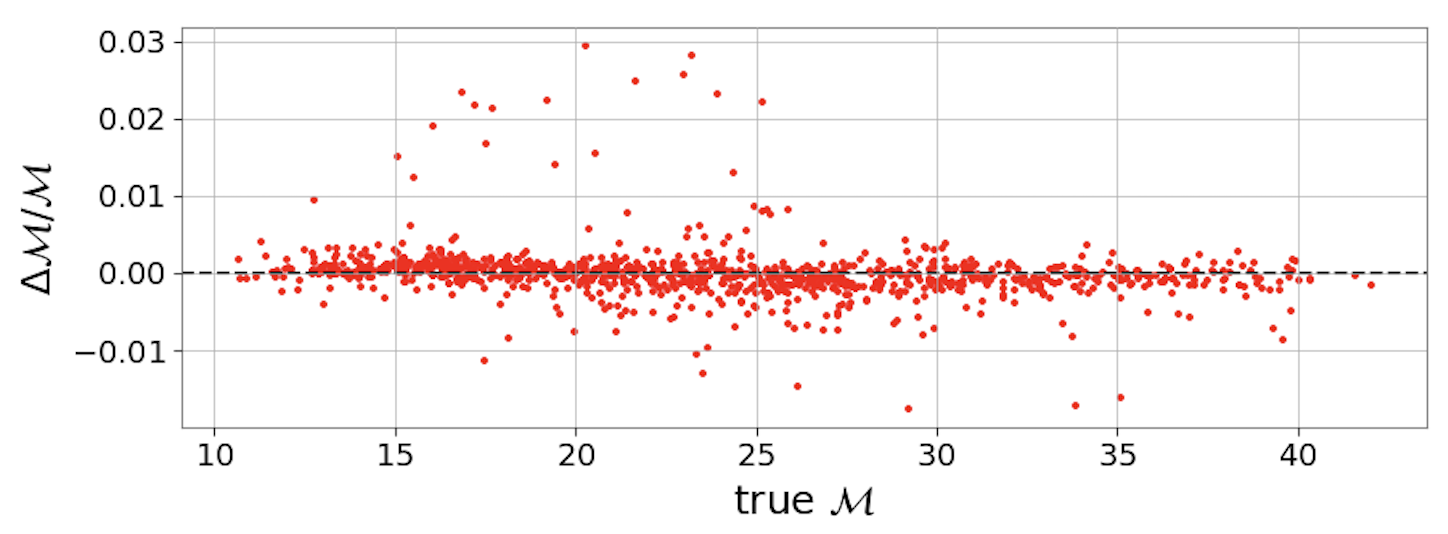}
     \end{subfigure}
     \begin{subfigure}{0.49\textwidth}
         \centering
         \includegraphics[width=\textwidth]{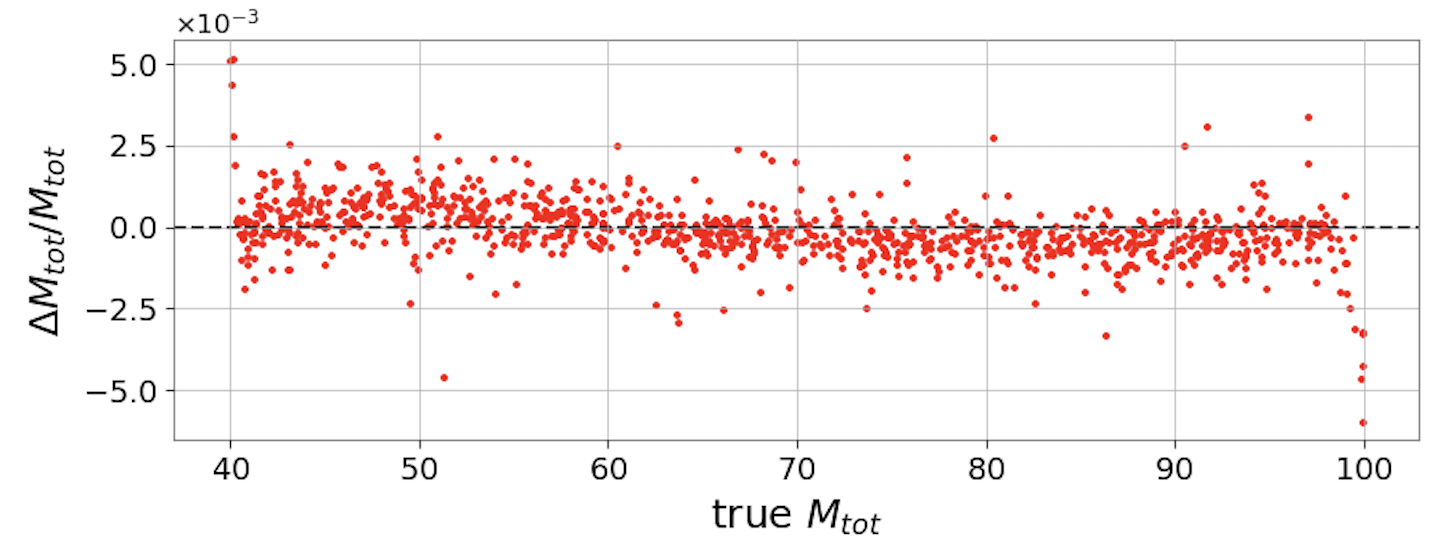}
     \end{subfigure}
     \hfill
     \begin{subfigure}[b]{0.49\textwidth}
         \centering
         \includegraphics[width=\textwidth]{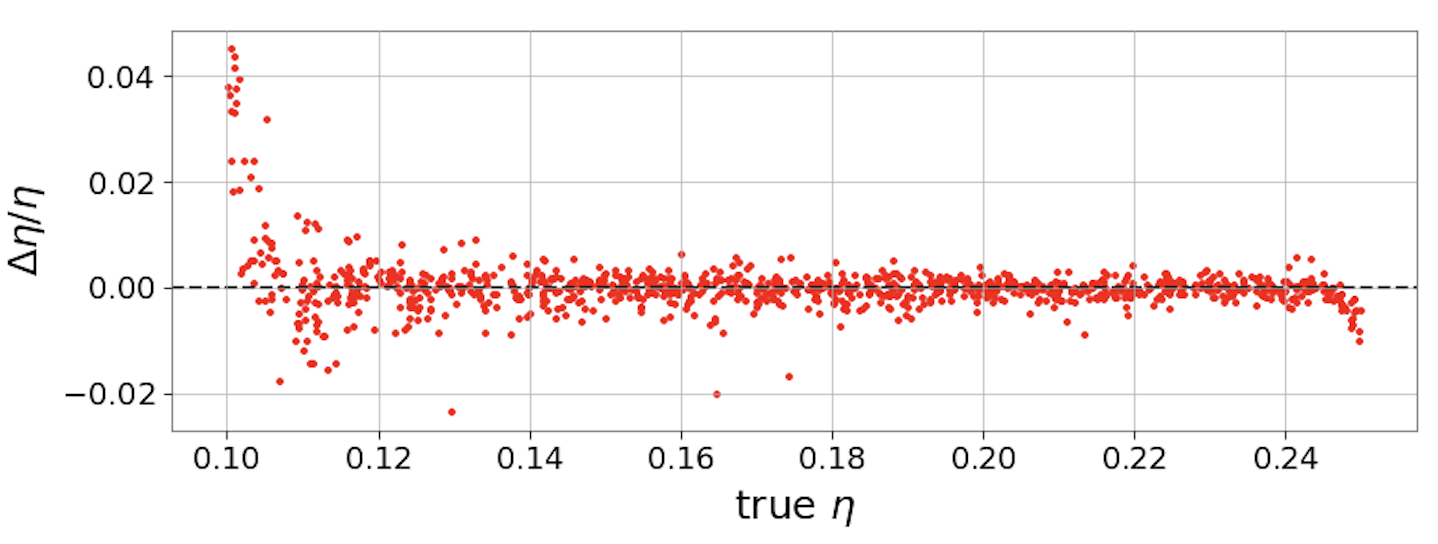}
     \end{subfigure}
     \begin{subfigure}{0.49\textwidth}
         \centering
         \includegraphics[width=\textwidth]{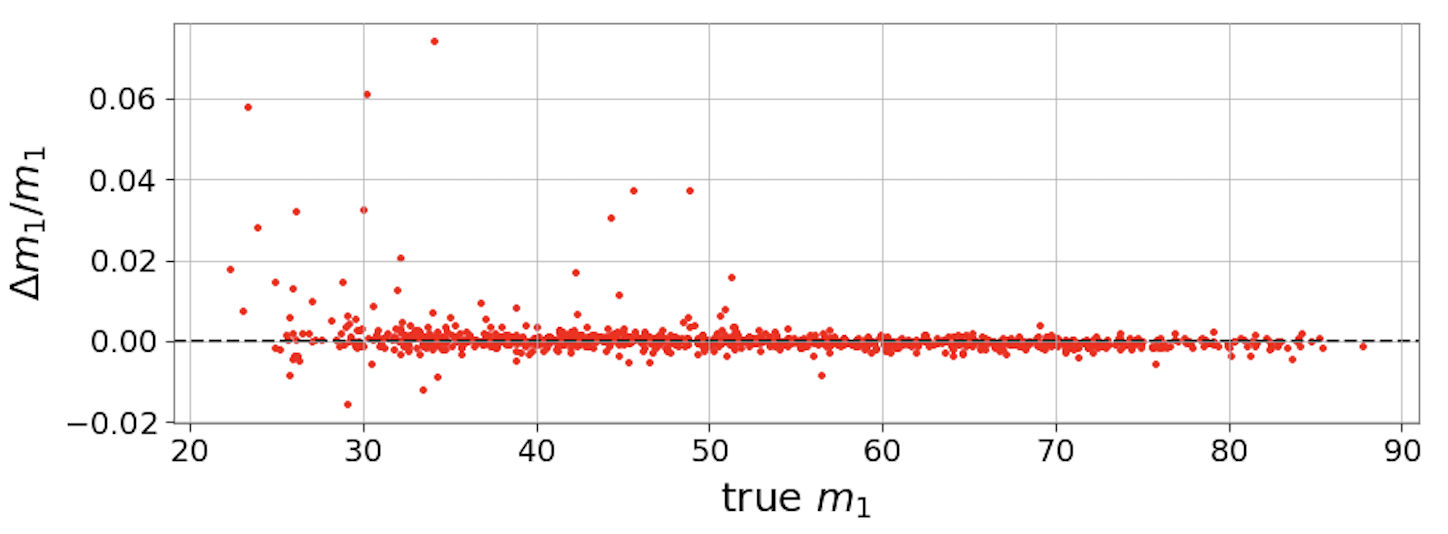}
     \end{subfigure}
     \begin{subfigure}{0.49\textwidth}
         \centering
         \includegraphics[width=\textwidth]{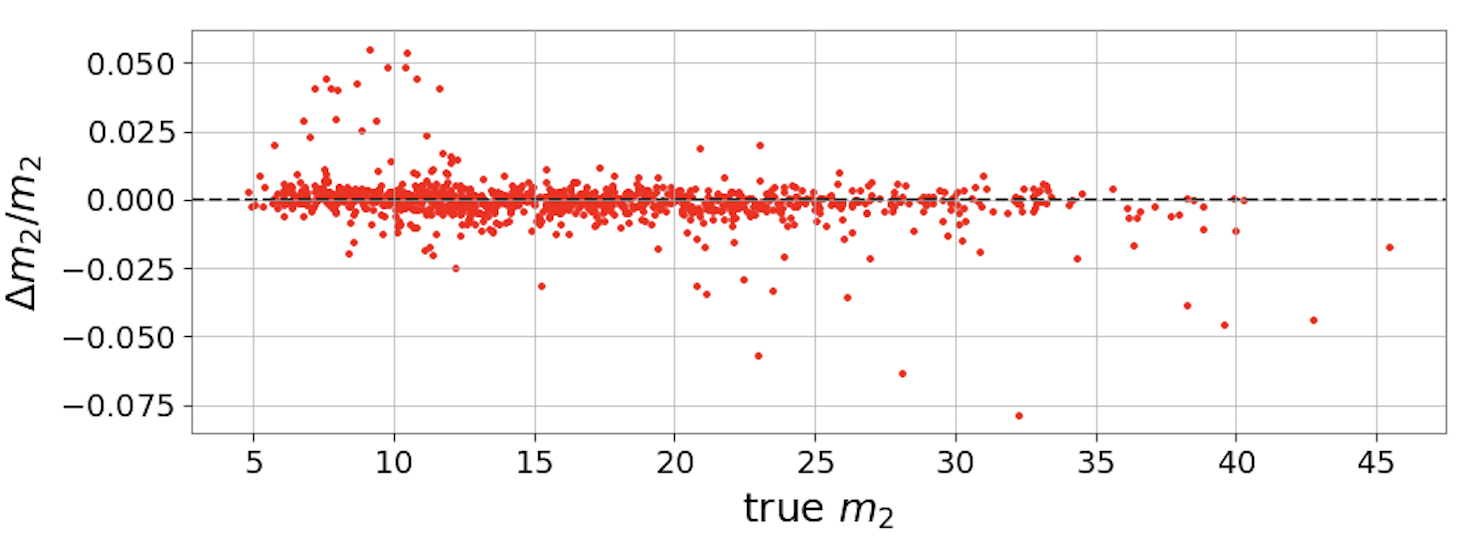}
     \end{subfigure}
    \caption{Relative errors of the inferred mass properties vs. true parameter. The black dashed lines represent zero relative error, while each red marker represents a signal from the test data.  }
    \label{fig:results}
\end{figure}

The performance of the network is tested by inputting the TF representations of signals from the test data set into the model, which infers a signal's mass properties within 2 ms. Fig. \ref{fig:results} presents the accuracy of the network via the relative errors for each of the inferred parameters. Looking first at the parameters directly output from the network, $\mathcal{M}$, $M_\mathrm{tot}$, and $\eta$ are inferred to within 3\%, 0.5\% and 5\% respectively. From these parameters, the component masses $m_1$ and $m_2$ are inferred (via Eqs. \ref{eq:masses}) to within 7\%. 
Looking more closely at the errors on the component masses, it can be seen that the majority of points lie within the 2\% error region, and that only a few outliers have errors up to 7\%. This is especially true for low mass values, the cause of which will be investigated in future work. In addition to this, the errors on $M_\mathrm{tot}$ and $\eta$ appear to follow a sigmoid trend, with increased errors towards the boundaries of the parameter ranges. We suspect that such behaviour arises from the sigmoid final layer in the network, and alternative functions will be investigated to overcome this problem.

There are many avenues to enhance this work, with the first steps to improve the accuracy and robustness of the network. This will be accomplished through the consideration of TF signals directly from cWB and through the exploration of a more detailed data set with increased parameter ranges. 
In addition to this, we plan to investigate the structure of our model further with the aim to decrease errors and biases across the parameter range. Furthermore, the natural progression of this project will be to consider further PN expansion terms in our governing equations, such as spin-orbit and spin-spin terms, which is required to generalise to spinning BBHs.

\section*{Acknowledgments}
This work acknowledges support from the European Union - Next Generation EU Mission 4 Component 1 CUP J53D23001550006 with the PRIN Project No. 202275HT58, and by ICSC – Centro Nazionale di Ricerca in High Performance Computing, Big Data and Quantum Computing, funded by European Union – NextGenerationEU. This research was partially supported by the Polish National Science Center OPUS grant no. 2021/43/B/ST9/01714, and by the National Aeronautics and Space Administration (NASA) under Award 80NSSC24K0767.

\bibliographystyle{ieeetr}
\bibliography{ref}

\end{document}